\documentclass[a4paper,11pt]{article}

\usepackage[a4paper,bindingoffset=0truemm,%
            left=20truemm,right=20truemm,top=30truemm,bottom=29truemm,
            footskip=0truemm]{geometry} 
\usepackage{multicol}
\usepackage{graphicx}
\usepackage{setspace}
\usepackage[version=3]{mhchem}
\usepackage[hang,flushmargin]{footmisc} 
\usepackage[psdextra]{hyperref}
\usepackage[
backend=biber,
sorting=none,
style=chem-acs
]{biblatex}

\newcommand{\abstracttitle}[1]{
\setstretch{1.15}	
 \begin{center}{\Large {\bf #1}}\end{center}
\setstretch{1.0}
}

\newcommand{\authors}[1]{
 \begin{center}{\bf #1} \end{center}
 \setstretch{1.0}
}

\newcommand{\addresses}[1]{
\setstretch{0.95} 
 \begin{center}{\small #1} \end{center}
\setstretch{1.0}
}

\newcommand{\Abstract}[1]{
 \begin{center}
 \begin{minipage}[t]{16cm}
   \setstretch{0.85}
   {\footnotesize {\bf Abstract} #1 }	
 \end{minipage}
 \end{center}
 \setstretch{1.0}
}

\newcommand{\writeto}[1]{
 \hspace*{-2.5mm} \footnote{\small E-mail: \href{mailto:#1}{#1}} 
  \hspace*{-3.0mm} 
}

\begin{document}
\abstracttitle{Assessing the dissociation hierarchy of aniline under UV-induced multiphoton ionization}

\authors{
S Muthuamirthambal$^1$, 
B Panja$^1$,
J Rejila$^1$,
R Sreeja $^1$,
M Theertha$^1$,
A Vishnumaya$^1$ and
U Kadhane$^{1}$\writeto{umeshk@iist.ac.in}
} 

\addresses{
$^1$Department of Physics, Indian Institute of Space Science and Technology, Kerala 695547, India
}

\Abstract{\hspace*{2mm}
The multiphoton ionization of the simplest aromatic amine, aniline, was examined utilizing a kinetic energy-correlated time-of-flight mass spectrometer at a wavelength of 266 nm. The primary and secondary fragment channels have been identified, and their plausible internal energy dependence has been discussed. Furthermore, the sequential loss fragment channel has been analyzed using the energy-correlated mass spectrum, revealing the fragment channels along with their respective parent mass exclusively. The computed energetics are corroborated with the experimental findings to further support the dissociation hierarchy.
} 

\section{Introduction}
Aniline (AN) is a simple aromatic amine commonly known as a precursor in the synthesis of complex molecules, including biomolecules~\cite{anjalin2020brief}. Due to its central importance from molecular physics to astrochemistry, extensive experimental and theoretical investigations have been done to understand its structure as well as its dissociation dynamics. The electronic structure and the excited state dynamics of neutral AN have been examined using various spectroscopic and computational techniques~\cite{kirkby2015comparing, roberts2012unraveling,  lykhin2021role, montero2011ultrafast, spesyvtsev2012shedding, selvaraj2023comprehensive,wojciechowski2003electronic}. The dissociation dynamics of AN in its cationic state have been studied under various energetic radiation conditions, including the ultraviolet (UV)~\cite{baer1982,zimmerman1990multiphoton, lubman1982mass}, vacuum ultraviolet (VUV)~\cite{selvaraj2023comprehensive, geng2020ionization, tseng2004photoisomerization, selvaraj2024fragmentation}, and electron impact (EI) ionization~\cite{le2002ionized,zeh2021cryogenic}. The primary fragmentation pathways of the AN cation are recognized as the loss of H and HNC; a detailed investigation on these channels, such as their appearance energy~\cite{selvaraj2023comprehensive}, decay rate~\cite{neusser1985unimolecular}, and the structure of the product ions~\cite{rap2022spectroscopic}, is being done. In contrast, the AN dication is prone to lose \ce{C_{n}H3^+} as its main fragmentation channels~\cite{selvaraj2024fragmentation}. These observations are further supported by quantum structural calculations and potential energy surface mapping~\cite{choe2009unimolecular}. In particular, the UV-induced multiphoton ionization study of aniline is driving more interest in characterizing the dissociation dynamics~\cite{beck2017intense, zimmerman1990multiphoton, baer1982}. Among several UV-induced MPI studies, Kuhledwind et al.~\cite{kuhlewind1985multiphoton} observed nine metastable dissociation channels using reflectron time-of-flight (ToF) mass spectrometry and discussed their dissociation sequence. Despite numerous investigations, a complete understanding of the aniline dissociation hierarchy under UV photoionization remains elusive. 

The present study revisited the AN dissociation process under the UV multiphoton ionization (MPI) condition using the high-resolution ToF mass spectrum, coupled with the parallel plate energy analyzer (PPA). The observed mass spectrum, together with the kinetic energy distribution of the ions, allows for the identification of both fragment ions and their originating parent ions. Furthermore, the mass spectrum of AN and \ce{AN-^{15}N} has been employed for a thorough examination of the compositions of dissociation product ions. In addition to the previously established dissociation channels, the current investigation reveals the sequential loss dissociation channels and their internal energy content, which have not been reported before. The ability to trace fragment ions in tandem with their parent ion allows for a definitive attribution of the primary and secondary fragment loss processes, which are extensively investigated and discussed. The calculated energetics of the potential dissociation routes are derived, and probable dissociation pathways are proposed. 

\section{Experimental and computational details}

\begin{figure}
    \centering
    \includegraphics[width=\linewidth]{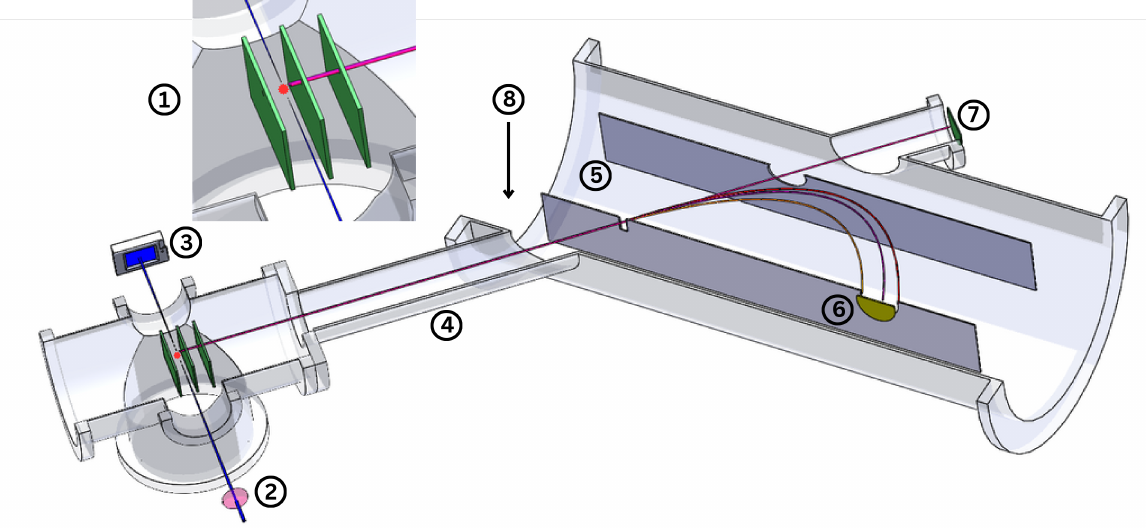}
    \caption{Schematic of the experimental setup. 1. expanded view of interaction region, 2. focusing lens, 3. energy meter, 4. drift region, 5. PPA, 6. ion PSD and 7. neutral PSD, 8. position where RGA mounted}
    \label{Schematic}
\end{figure}

The experiment was carried out at the Atomic and Molecular Physics (AMP) laboratory, IIST, India, using the high-resolution ToF mass spectrometer equipped with the parallel plate energy analyzer (PPA. The schematic of the experimental setup is shown in Fig.~\ref{Schematic}. The detailed information on the experimental setup can be found elsewhere~\cite{vinitha2019high,muthuamirthambal2025intracluster}; only the necessary details are provided here. A commercial liquid aniline sample was used at room temperature without further purification. The target molecule was injected through a simple hypodermic needle, and the effusive jet intersected with the focused laser beam in the interaction region of the spectrometer. The ionized molecules are then extracted and accelerated in the fields of 43 V/mm and 316 V/mm, respectively, towards the field-free region (FFR) and passed through a PPA and splat on the position sensitive detector (PSD). The raw signal from the PSD was amplified and discriminated, then sent to the time-to-digital converter (TDC) for the ToF and position measurements. The laser with the 6ns pulse width Nd: YAG laser was used for the multiphoton study, the energy per pulse was maintained at 20-100$\mu$J, and the focused laser beam was 200$\mu$m in diameter. The maximum pressure on the interaction chamber during the experiment was \ce{3 $\times$ 10^{-7}} mbar and the PPA chamber pressure was maintained at \ce{4 $\times$ 10^{-8}} mbar. 

The fragmentation channel and its corresponding parent ion are identified from the energy correlation ToF mass spectrum analysis as follows. The statistical dissociation process adheres to an Arrhenius decay, resulting in probabilistic dissociation from the laser site where they ionized until the ion reaches the PSD. If ions dissociate in the extraction or acceleration region, their energy will vary; however, if they dissociate within the FFR, the fragment ions will possess the same velocity as the parent ion, but their kinetic energy will differ based on their mass.  The ions produced by decay inside the FFR can be captured by adjusting the PPA bias voltage. In this way, the fragment ions can be recorded at the same ToF as their parent but with a different kinetic energy proportional to their mass if the daughter ions are made to follow the exact same trajectory as the originating parent ions inside PPA. Finally, with the energy and mass conservation principles, the energy of the fragment ions measured for the set PPA bias voltage and its parent ions mass obtained from its ToF allows us to calculate the mass of the daughter ion fragment. 

Density functional theory calculations were performed using the GAUSSIAN 16 software~\cite{g16} to obtain the energetics of the possible dissociation pathways. The Potential energy surface (PES) scans were conducted at the B3LYP/6-311++ G(d,p) theoretical level to identify the structures of the stationary points involved. Geometries were optimized at an equivalent level, and vibrational frequencies were used to characterize the structures as minima (intermediates) and first-order saddle points (transition states). Intrinsic reaction coordinate (IRC) computations were performed for critical steps to confirm that the associated transition state indeed connects the local minima on both sides. 

\section{Results}
\subsection{Multiphoton ionization mass spectrum}
\begin{figure}
    \centering
    \includegraphics[width=0.8\linewidth]{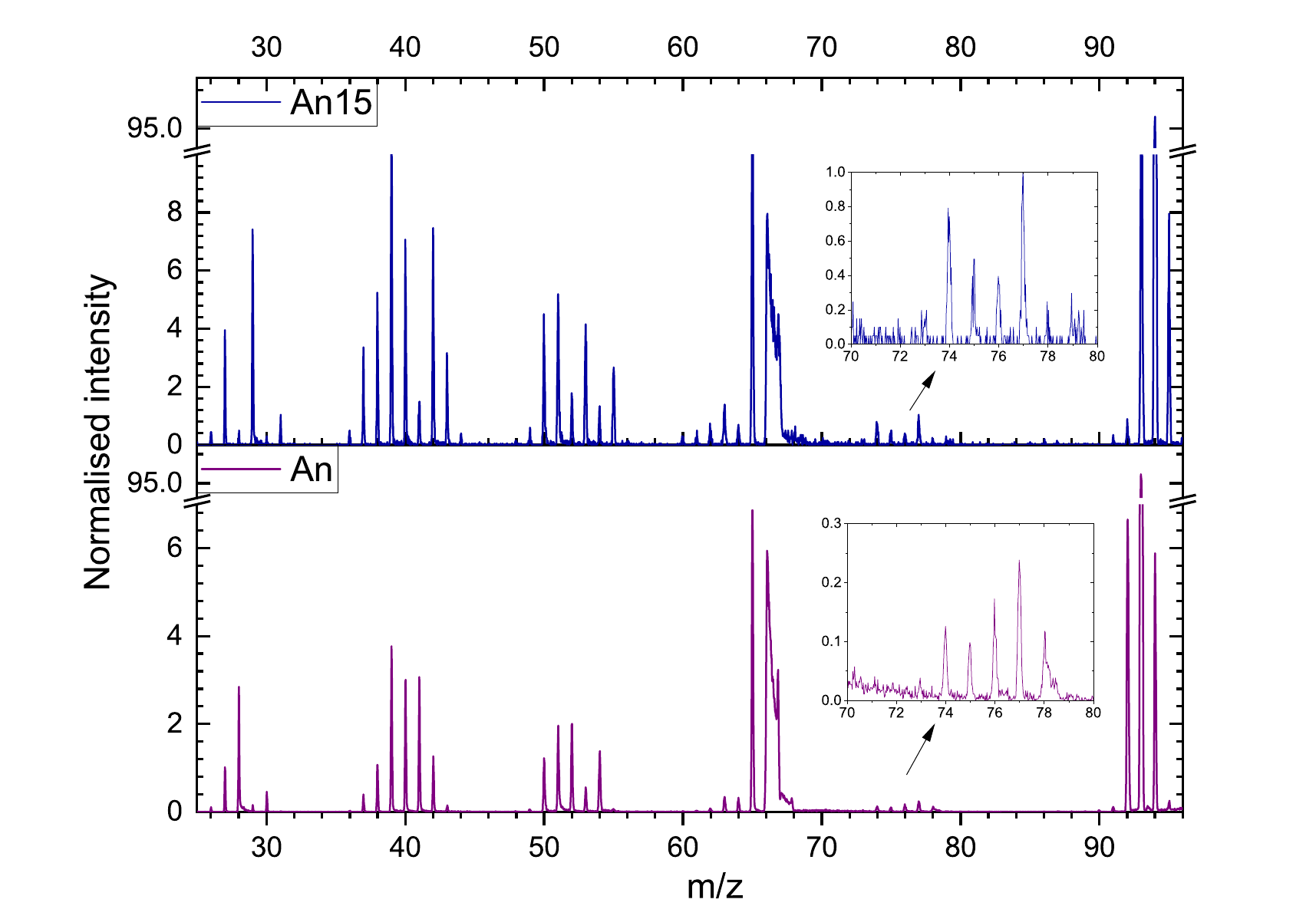}
    \caption{The multiphoton ionization mass spectrum of AN and \ce{AN-^{15}N} at 266 nm photon wavelength. The mass spectrum is normalised to the parent ion yield. The inset shows the expanded view of the peaks between 70 and 80.}
    \label{ToF}
\end{figure}

The multiphoton ionization mass spectra of AN and nitrogen-labeled aniline (\ce{AN-^{15}N}) were acquired using a 266 nm wavelength photon beam under identical experimental conditions. Fig.~\ref{ToF} illustrates the measured mass spectrum of AN and \ce{AN-^{15}N}, with the mass spectrum normalized to the yield of the respective parent ion. The mass spectrum shows the typical characteristics of an AN fragmentation pattern as reported in the previous studies ~\cite{kuhlewind1985multiphoton,zimmerman1990multiphoton,selvaraj2023comprehensive}. The AN parent ion (\ce{C6H5NH2^+}) constitutes the dominant peak at m/z 93. The most intense fragment ion is observed at m/z 66, signifying the neutral loss of HNC~\cite{selvaraj2023comprehensive,zimmerman1990multiphoton,baer1982,choe2009unimolecular}, resulting in the product ion \ce{C5H6^+} with a relative yield of 36\% . The broadening and elongated tail of this channel up to m/z 72 vividly highlights the metastable dissociation process of the HNC loss channel. The second similarly intense fragment ions at m/z 92 and 65 correspond to the loss of H and HCNH/HNC+H, yielding roughly 8.5\%. The fragment channels above m/z 70 exhibit moderate intensity, whilst the remaining fragment peaks, particularly within the ranges of 50 to 55 and 37 to 41, have yields of 4\% to 5\%. The peak at m/z 28, expected to be \ce{HCNH^+}, yields around 3\% and possesses a metastable decay characteristic. Many fragment channels exhibit characteristics of metastable feature, which will be discussed in the following section.

\subsection{Energy correlated mass spectrum}
\begin{figure}
    \centering
    \includegraphics[width=\linewidth]{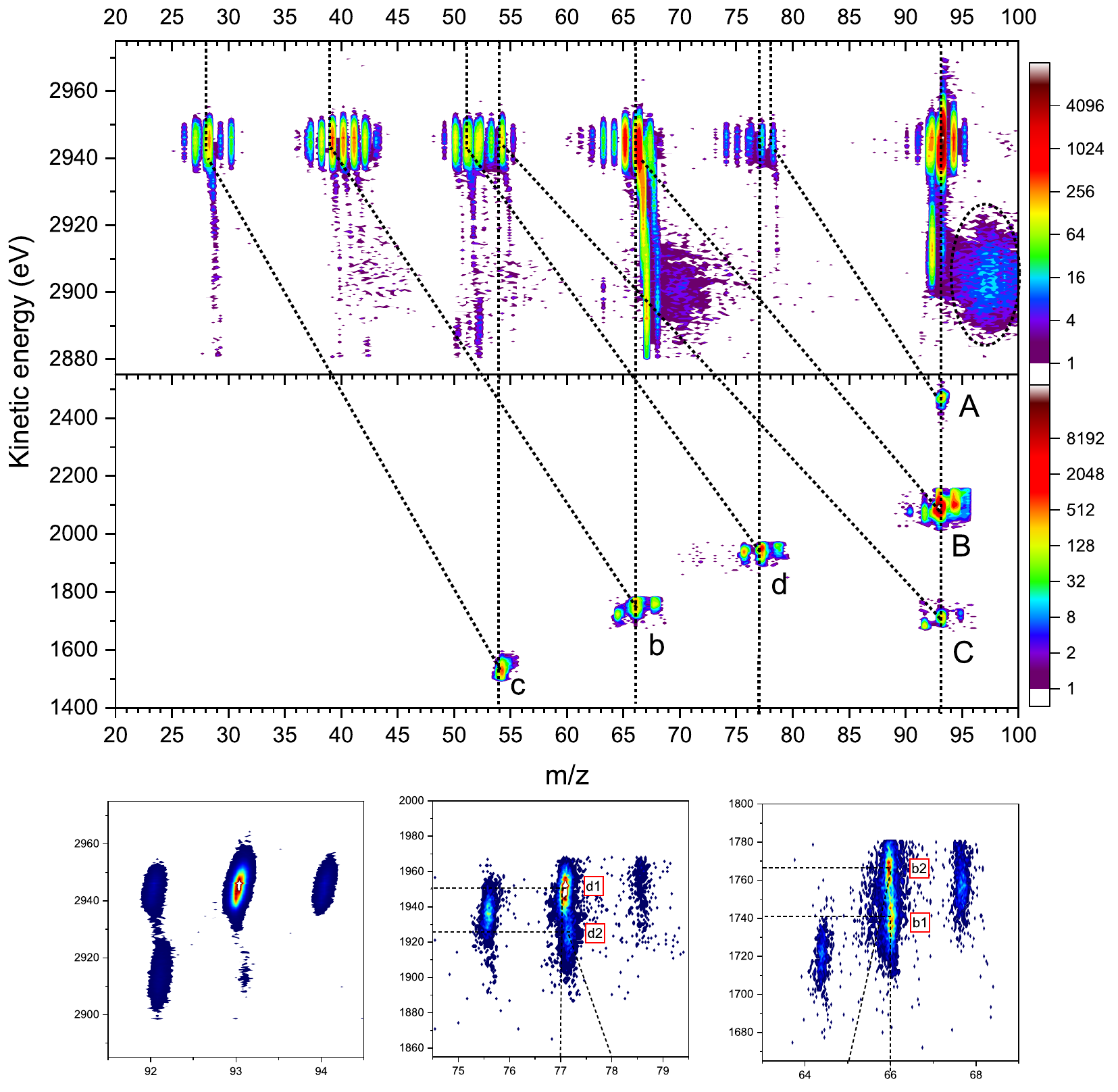}
    \caption{The energy and mass correlation plot for AN recorded at different PPA set voltages. The dotted circle shows the ions which are formed due to the charge exchange process between the AN parent ions and the background neutral molecules. The three expanded panels show the detailed view of the ions \ce{C6H6N^+} (m/z 92), \ce{C4H3^+} (m/z 51) and \ce{C3H3+} (m/z 39) formed inside the FFR due to the slow dissociation process. }
    \label{2D}
\end{figure}

The energy-correlated ToF mass spectrum of AN is shown in Fig.~\ref{2D}. The upper panel with the kinetic energy range of 2880 to 2970 eV presents the mass spectrum of ions produced in the interaction region, with their kinetic energy distribution. The lower panel, in the range of 1400 to 2500 eV, shows the fragment ions formed in the FFR, as recorded by the appropriate PPA bias voltages. The dotted lines connect the ions formed in the FFR due to the slow decay process with their originating parent ions. The islands \textbf{A}, \textbf{B}, and \textbf{C} exhibit the primary fragment channels at m/z 78, 66, and 54, derived from the AN parent ion. The islands \ce{b}, \textbf{c}, and \textbf{d}, corresponding to the sequential neutral loss, originated from the primary fragment channel ions. The expanded image illustrates the H loss channel alongside the distribution of the parent ions (left). The prompt H loss channel exhibits an energy distribution akin to that of the parent ions, while the H loss occurring within the FFR demonstrates an energy difference corresponding to H loss. The two parent ions at m/z 77 (\textbf{d1}) and 78 (\textbf{d2}) serve as the precursors for the product ion at m/z 51, resulting from the loss of neutral masses of 27 and 26, respectively (middle). Likewise, the ions corresponding to m/z 39, resulting from the neutral loss of 27 and 26 from the primary fragment ions at m/z 66 (\textbf{d1}) and 65 (\textbf{d2}), respectively (right). 

\subsection{Fragment channel composition}
The comparison of the mass spectra of AN with \ce{AN-^{15}N} aids in identifying the nitrogen-containing fragment channels ions. The mass transition from 78 to 79 in the \ce{AN-^{15}N} spectra indicates that 50 to 60\% of the channel at m/z 78 includes nitrogen, while the remaining portion does not. Thus, the channel at m/z 78 may result from the neutral loss of NH or \ce{CH3} from the aniline parent ion and lead to the formation of the ions \ce{C6H6^+} as well as \ce{C5H4N^+}, respectively. The lack of a shift in the fragment channel at the m/z 77 peak signifies a nearly total depletion of \ce{NH2} associated with this channel and formation of the product ion \ce{C6H5^+}. The constant yield of peaks at m/z 66 and 65 indicates that the elimination of HNC and H+HNC or HNCH has been associated with these channels, producing the product ions \ce{C5H6^+} and \ce{C5H5}, respectively. For the fragment channel at m/z 54, there exists a $>$90\% possibility that the ions comprise nitrogen, which is \ce{C3H4N^+} possibly after losing neutral \ce{C3H3}. The ions at m/z 51 and 39 are entirely assigned to \ce{C4H3^+} and \ce{C3H3}, respectively, whereas m/z 28 is ascribed to \ce{HCNH^+} ions.  

\section{Discussions}
The MPI mass spectrum of AN presented in this study predominantly features the parent ion alongside fragment ion channels, consistent with previous findings~\cite{selvaraj2023comprehensive,kuhlewind1985multiphoton,zimmerman1990multiphoton}. The electronic excited state and lifetime of AN have been extensively studied~\cite{spesyvtsev2012shedding,spesyvtsev2012ultrafast,roberts2012unraveling,thompson2013following,kirkby2015comparing,montero2011ultrafast}. The absorption of a single photon excites the AN molecule to its long-lived ($>$1ns) first singlet state (S1 $\pi \pi^*$), which likely relaxes to S0~\cite{spesyvtsev2012shedding,roberts2012unraveling,montero2011ultrafast,thompson2013following}. It has been suggested that, due to the slight geometric variations between the S1 and S0 states, the vertical ionization from the S0 state may be about equivalent to that from the S1 state~\cite{moll1984time}. The present study utilizes a nanosecond pulsed laser, expecting that second photon absorption occurs via the S1 state, ultimately attaining the ionization continuum, with an ionization potential of 7.7 eV~\cite{montero2011ultrafast,selvaraj2023comprehensive}. The ladder-switching process is possibly accountable for the production of parent ions and fragment ions that absorb additional photons for further dissociation. The focus is given on investigating significant primary and sequential loss channels that exhibit metastable characteristics, including previously undocumented fragment channels.  

\begin{figure}
    \centering
    \includegraphics[width=0.7\linewidth]{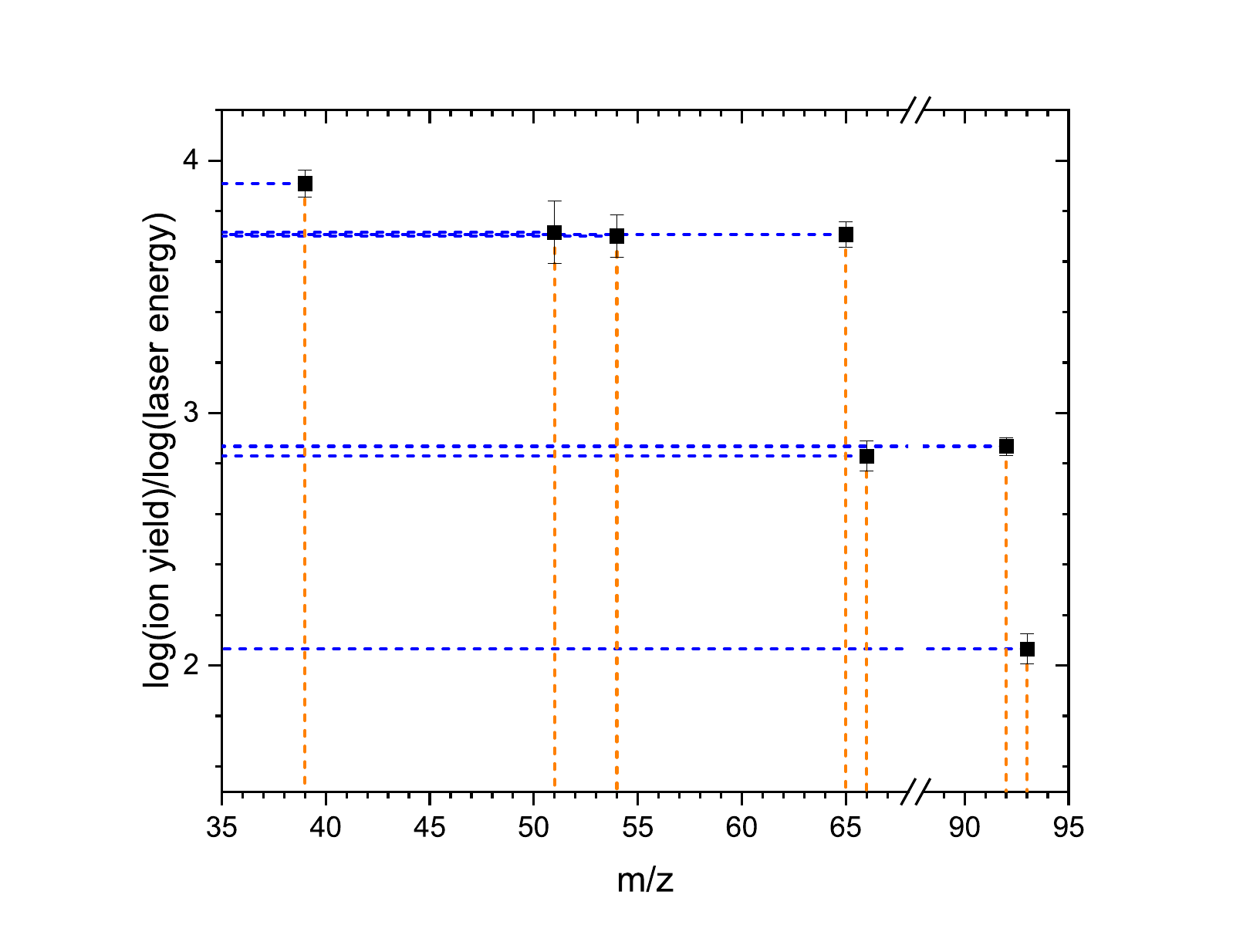}
    \caption{The laser power dependency for the selected masses in the multiphoton ionization at 266nm.}
    \label{slopes}
\end{figure}

The laser pulse energy dependence on the relevant yield of ions is determined, and the results of the photon number dependence of these masses are presented in Fig.~\ref{slopes}, which is consistent with the previous investigations~\cite{kuhlewind1985multiphoton,zimmerman1990multiphoton}. The parent ion demonstrates a two-photon dependence, whereas the fragment ions at m/z 66 and 92 exhibit a three-photon dependence. The ions at m/z 65, 54, 51, and 39 have a four-photon dependence, indicating the substantial energy necessary for the dissociation process. The calculated possible low-energy dissociation pathways are compared with the experimental evidence and evaluated herein. The Fig.~\ref{Seq} depicts the structure and the computed maximum energy demand for the primary and dissociation routes, with all energies referenced to the neutral AN structure.  

\subsection{ \texorpdfstring{\ce{CH3}}{} loss followed by HNC loss}
\begin{figure}
    \centering
    \includegraphics[width=\linewidth]{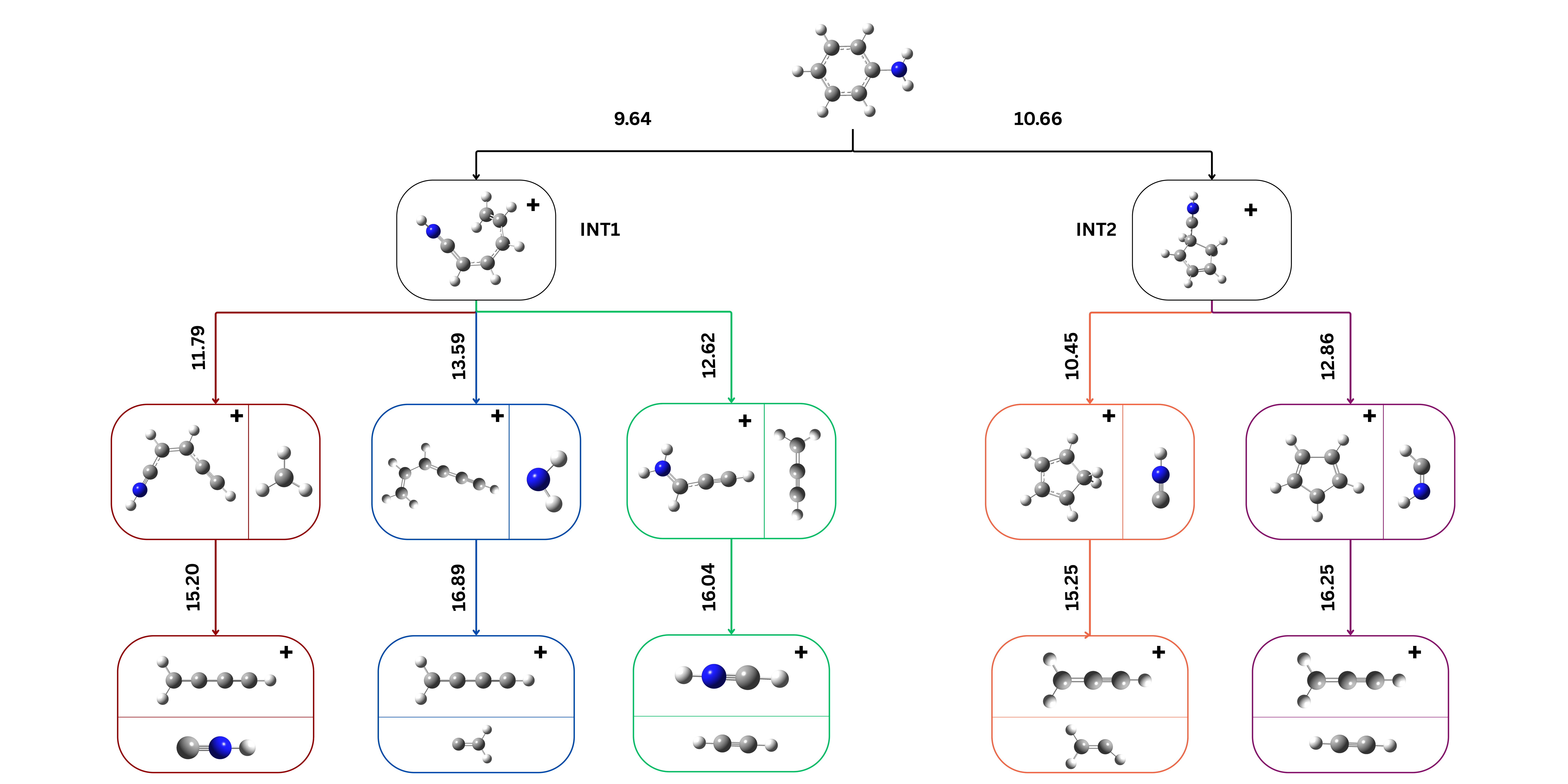}
    \caption{The dissociation hierarchy of the AN under MPI observed in the present work. The calculated maximum energy requirements are given in eV with respect to the AN neutral structure.}
    \label{Seq}
\end{figure}

The metastable dissociation of the fragment ion at m/z 78, depicted in island \textbf{A} in Fig.~\ref{2D}, corresponds to the neutral loss of \ce{CH3} or NH from the AN parent ion, which was not observed in previous studies. The comparison of the mass spectra for AN and \ce{AN-^{15}N} indicates that the peak at m/z 78 likely comprises $>$50\% nitrogen-containing compound \ce{C5H4N^+} ions and remaining \ce{C6H6^+} ions. However, the ensuing dissociation product ion at m/z 51 lacks nitrogen; consequently, the elimination of \ce{CH3} from the parent ion produces \ce{C5H4N^+} ions, which may further dissociate by the loss of HNC or NH from the parent ion, resulting in \ce{C4H3^+} ions through the release of \ce{C2H3} as a neutral fragment. The probable dissociation pathways are subsequently calculated for the elimination of \ce{CH3} via a structure \textbf{INT1} in Fig~\ref{Seq}, necessitating an energy of 11.79 eV. The \ce{C5H4N^+} product ion further dissociates into the \ce{C4H3^+} ion by releasing neutral HNC, requiring an energy input of 15.20 eV compared to the neutral AN structure. Conversely, the primary dissociation of neutral NH from the parent ion (m/z 78) occurs via the same \textbf{INT1} structure, necessitating a minimum of 16.13 eV, while the subsequent neutral loss of \ce{C2H3} demands 20.61 eV, significantly exceeding the loss of HNC followed by the \ce{CH3} process. The probability of direct cleavage of NH from the parent ring structure also results in a greater energy need, 15.92 eV. Therefore, the four-photon dependence of the channel at m/z 51 indicates that the loss of NH could serve as the primary dissociation pathway, which is not likely to participate in subsequent neutral losses. Thus, the \ce{CH3} elimination pathway is effective for yielding the product ion at m/z 51 through sequential dissociation via \textbf{INT1}.  

\subsection{\texorpdfstring{\ce{NH2}}{} loss followed by \texorpdfstring{\ce{C2H2}}{} loss}
The fragment ion at m/z 77, \ce{C6H5^+}, corresponds to the loss of \ce{NH2} from the AN parent ion, determined by the \ce{AN-^{15}N} mass spectrum. The fragment ion \ce{C6H5^+} subsequently loses a neutral \ce{C2H2}, resulting in the formation of a secondary product ion at m/z 51, which is \ce{C4H3+} as described in the previous studies~\cite{kuhlewind1985multiphoton,selvaraj2023comprehensive}. The common \textbf{INT1} structure contributes to the \ce{NH2} scission process, necessitating an energy input of 13.59 eV, thereafter decaying into the product ion \ce{C4H3^+}, which requires a maximal energy of 16.89 eV. The absence of the metastable decay signature of m/z 77 from the AN parent ion indicates that this channel has a faster dissociation rate, recommending that the internal energy exceeds the dissociation energy threshold. Therefore, the total absorption of four photons (18.64 eV) must have contributed to this channel, which is also adequate for causing the sequential loss of \ce{C2H2} (16.89 eV). The finding of the onset energy of 15 eV of this channel in the VUV-induced dissociation study~\cite{selvaraj2023comprehensive} for the channel at m/z 77 further corroborates the suggested energetics and dissociation sequence.   

\subsection{HNC loss followed by \texorpdfstring{\ce{C2H3}}{} loss}
The most intense primary fragmentation channel at m/z 66 yields the product ion \ce{C5H6^+},  cyclopentadiene cation (\ce{CP^+}), which arises from the neutral loss of HNC, having a significant metastable feature observed on island \textbf{b1}.  The dissociation mechanism of this channel has been extensively investigated in previous studies\cite{selvaraj2023comprehensive,choe2009unimolecular, lifshitz1983time, baer1982}. This channel is likely to occur through the five-membered intermediate structure (\textbf{INT2}) with an energy requirement of 10.66 eV. Although the \textbf{INT2} structure requires roughly 1 eV more than INT1, if the HNC loss occurs through \textbf{INT1}, a maximum of 12.48 eV is necessary in the subsequent transition state steps. Therefore, in line with previously reported studies, the HNC loss unambiguously occurs through the \textbf{INT2}. The preliminary product ion at m/z 66 undergoes a consecutive neutral loss of mass 27, resulting in the product ion at m/z 39, as illustrated in Fig.~\ref{2D}. The dissociation of the cyclopentadiene ion, yielding the secondary product ion \ce{C3H3^+} after the neutral loss of \ce{C2H3}, occurs at an energy of 15.25 eV following successive ring opening~\cite{kuhlewind1985multiphoton}. 

\subsection{HNCH loss followed by \texorpdfstring{\ce{C2H2}}{} loss}
The primary fragment ion at m/z 65, identified as \ce{C5H5^+}, can be ascribed to the elimination of HNC+H or the intact HNCH from the AN parent ion following the absorption of four photons~\cite{selvaraj2023comprehensive,kuhlewind1985multiphoton}. Nonetheless, the trace of \ce{C5H5^+} within FFR, resulting from the hydrogen removal from \ce{C5H6}, is absent in the Fig.~\ref{2D} and also the sharp feature of the peak in the mass spectrum in Fig.~\ref{ToF}. This suggests two possibilities: that the \ce{C5H5^+} ion is formed directly from the AN parent ion by the loss of intact HNCH, indicating that \ce{C5H5^+} may not be derived from \ce{C5H6} due to hydrogen loss, or that \ce{C5H6^+} dissociates rapidly into \ce{C5H5^+} due to significant internal energy exceeding the dissociation energy threshold required to lose neutral hydrogen. The distribution of ions near island \textbf{B} (to the left) confirms the formation of \ce{C5H5}, which dissociates from the parent AN ion within the FFR. The calculated low-energy dissociation pathway for the HNCH loss from the AN parent ion via \textbf{INT2} requires an energy of 12.86 eV. Whereas, the onset energy of this channel in the VUV photodissociation investigation is 13.5 eV, while the calculated energy for the sequential loss of HNCH and H is approximately 14.5 eV, which contradicts this sequential HNC+H dissociation possibility~\cite{selvaraj2023comprehensive}. The four-photon (18.64 eV) dependence of this channel suggests that there will be around 5.8 eV of extra energy available within the molecule following the intact HNCH loss, which may cause both the faster dissociation and the secondary dissociation channels. Thus, it is possible that the channel m/z 65, cyclopentadienide ion, corresponds to the mostly intact loss of HNCH from the aniline parent ion through \textbf{INT2}. It should be noted that, the multiphoton study carried out by Kuhlewind et. al.~\cite{kuhlewind1985multiphoton}, has proposed HNC+H loss possiblity from the metastable ion spectrum performed using the reflectron ToF technique. From island \textbf{b2} in Fig~\ref{2D}, it is evident that the primary fragment ion \ce{C5H5} subsequently decomposes into the secondary product ion at m/z 39, the \ce{C3H3^+} ion, by expelling neutral \ce{C2H2}. The cyclopentadienide ion undergoes hydrogen migration within the ring, resulting in the opening of the ring structure, the loss of \ce{C2H2}, and the formation of \ce{C3H3^+}, necessitating an energy of 16.25 eV. Given the sequential dissociation pathway with an energy need of 16.25 eV, the residual energy of around 2 eV after four photon absorptions justifies the slow decay process. 

\subsection{\texorpdfstring{\ce{C3H3}}{} loss followed by \texorpdfstring{\ce{C2H2}}{} loss}
The fragment channel at m/z 54 consists of more than 90\% nitrogen-containing compounds, specifically \ce{C3H4N^+} ions, resulting from the decomposition of the aniline parent ion by the loss of a \ce{C3H3} neutral fragment. The island C in Fig.~\ref{2D} corroborates the origins of this channel from the aniline parent ion.  The dissociation pathway can proceed through the \textbf{INT1} structure via hydrogen migration steps, culminating in the elimination of neutral \ce{C3H3}, with an energy requirement of 12.62 eV, far lower than the total energy of four photons. The three-photon absorption may activate this dissociation channel, leading to metastable dissociation. The four-photon absorption can impart a substantial internal energy of around 6 eV above the dissociation threshold, resulting in fast dissociation, as evidenced by the sharp peak in the mass spectrum. However, considering the substantial dissociation that occurs, allocating roughly 1 eV for the decomposition of the \ce{C4H3N^+} parent ion of AN, the remaining 5 eV is adequate to activate the subsequent dissociation process from this product ion. The estimated findings depicted in the Fig.~\ref{Seq} suggest the energy difference of approximately 3.5 eV between primary and secondary dissociation is well within the excess energy of 5 eV. Therefore, it is plausible that the absorption of four photons can efficiently induce both the major dissociation channel at m/z 54 and the secondary breakdown products \ce{HNCH^+} and \ce{C2H2} from the \ce{C3H4N^+} ion. 
The dissociation channel at m/z 54 in the electron impact and multiphoton ionization spectrum of aniline is exclusively ascribed to the nitrogen-containing molecular ion~\cite{kuhlewind1985multiphoton,zimmerman1990multiphoton}. Nevertheless, the present analysis identified around 90\% of nitrogen-containing molecules for this channel, which is consistent. The peak at m/z 53 may correspond to a nitrogen component, possibly shifted to m/z 54 in the \ce{AN-^{15}N} mass spectrum, which has not been considered in other studies. The removal of the neutral nitrogen component, HCCN, as described in the prior study on VUV-induced photodissociation~\cite{selvaraj2023comprehensive}, may not significantly influence the dissociation pathway at m/z 54.

\section{Conclusion}
Extensive experimental investigations of aniline under UV-induced MPI conditions have been conducted, and the dissociation hierarchy is delivered. The comparative mass spectra between AN and \ce{AN-^{15}N} reveal the distinct fragmentation channels involving nitrogen-containing species. The dominant dissociation channels and their corresponding parent ions are assigned from the observations gained from energy-correlated ToF mass spectrum measurements. Furthermore, the temporal progression of the dissociation process was characterized, highlighting a sequential fragmentation trend. The calculated PES supports the experimental evidence and offers insight into competing dissociation mechanisms. 

\printbibliography
\end{document}